# Parametric resonance magnetometer based on elliptically polarized light yielding three-axis measurement with isotropic sensitivity

Gwenael Le Gal,[1,2] Laure-Line Rouve,[2] Agustin Palacios-Laloy[1,*]

[1]CEA-LETI, Minatec Campus, F-38054 Grenoble, France. Univ. Grenoble Alpes, F-38000 Grenoble, France
[2]University Grenoble Alpes, CNRS, Grenoble INP, G2Elab, 38000 Grenoble, France
*Corresponding author: Agustin.palacioslaloy@cea.fr

**We present here a new parametric resonance magnetometer scheme based on elliptically polarized pumping light and two radio-frequency fields applied along the two optical pumping directions. At optimum ellipticity and radio-frequency fields amplitudes the three components of the magnetic field are measured with an isotropic sensitivity. Compared to the usual alignment-based parametric resonance magnetometers, the sensitivity is degraded by a factor 2 for two components of the magnetic field but improved by a factor 11 for the third one. This new magnetometer configuration could be particularly interesting for geophysics and biomedical imaging.**

Currently, optically pumped magnetometers (OPMs) can reach excellent sensitivities similar to SQUIDs [1] but without requiring cryogenics. This opens new prospects to precisely measure magnetic fields in space exploration [2], geophysics [3], studies of fundamental symmetries [4], magnetic imaging of biological currents in magnetocardiography (MCG) [5,6], and magnetoencephalography (MEG) [7,8]. Both for MEG and MCG, some recent studies suggest that a tri-axial magnetometer would improve the accuracy of source reconstruction [9,10], as well as the noise rejection [11], as far as the sensitivity is isotropic [10].

Such an isotropic sensitivity is not straightforward for OPMs, due to the symmetry breaking by the polarization of the pumping light. For instance magnetometers based on the Hanle effect allow a real-time vector measurement of the components of the magnetic field [1,12] which are orthogonal to the characteristic direction of pumping. In order to use a single light beam, parametric resonances magnetometers (PRM) are often used instead [13]. Fig. 1(a) shows a typical PRM based on alkali atoms [14–16], optically pumped toward an oriented state, i.e. a state with $\langle S_k \rangle \neq 0$ where $\vec{k}$ is the propagation direction of the circularly polarized pump beam. Two orthogonal radio frequency (RF) fields allow to measure the two components of the magnetic field parallel to them [14,17–19]. Unlike the Hanle effect magnetometer, the third component, parallel to $\vec{k}$, can be measured thanks to non-secular terms [17] but with a sensitivity much worse than the others.

An atomic state with F > 1/2 can be pumped using circularly polarized light toward an oriented state, or, using linearly polarized light, toward an aligned state (i.e. a state with $\langle 3S_e^2 - \vec{S}^2 \rangle \neq 0$ where $\vec{e}$ is the direction of the pump-light electric field $\vec{E}_0$ [20]). Using alignment instead of orientation in a PRM (Fig. 1(b)) yields a better sensitivity for the worst resolved axis [18], which in this case is parallel to $\vec{e}$, relatively to the two other axes.

Such PRMs based on the $2^3S_1$ [4]He metastable state (F = S = 1) proved their ability to detect biomagnetic signals both in MCG [6] and MEG [8].

We introduce here a new [4]He PRM configuration based on elliptically polarized pumping light that delivers vector measurement of the three components of the magnetic field with isotropic sensitivity. Indeed using elliptically polarized pumping light one can simultaneously prepare both an orientation and an alignment in the atomic ensemble, with a ratio fixed by the light ellipticity. Since $\vec{e}$ and $\vec{k}$ are orthogonal, and since the evolutions of orientation and alignment within a magnetic field are decoupled [18], we studied if their combination yields well resolved measurements of the three components of the magnetic field.

As a starting point we studied Hanle effect resonances as a function of the light ellipticity. The sensitivity of a PRM is proportional to $A = a/\Gamma^2$ [17], where $\Gamma$ is the half-width-half-maximum (HWHM) of the Hanle resonance and $a/\Gamma$ is its amplitude. We investigated experimentally the variation of $A$ as a function of the light ellipticity. The experimental setup for doing so is shown in Fig. 1(d). It comprises a 1-cm diameter and 1-cm length cylindrical cell filled with 9-Torr high purity helium-4, where the metastable state is populated using a high-frequency (HF) capacitively coupled discharge. An external cavity laser diode (Sacher Cheetah TEC 50) generates the pump beam. To keep it tuned with [4]He $D_0$ transition (at 1083.205 nm) a wavelength-meter (HighFiness WS-7) locks the laser diode temperature. The laser is collimated to 7-mm waist, and goes through a linear polarizer with transmission axis set at an angle $\varphi$ from the $\vec{y}$ axis, and a $\lambda/4$ zero-order waveplate (Thorlabs WPQ10M-1064), with its fast axis parallel to the $\vec{y}$ axis. The optical power is set to ~250 µW at cell input. The helium cell is placed inside two sets of triaxial coils: the inner one generating the RF fields, the outer one generating the magnetic field sweeps. The cell and coils are put inside a five-layer µ-metal magnetic shield. After crossing the cell, a lens focuses the light on an In-Ga-As photodiode, connected to a homemade transimpedance amplifier with $R_T = 23.8$ kΩ. We measure the DC photodetection signal while sweeping sequentially each component of the magnetic field of ±300 nT, with the others set to zero. Here no RF fields are applied. Fig. 1(e) shows the experimental results, along with the theoritical prediction calculated within the so-called "three-step approach" [21].

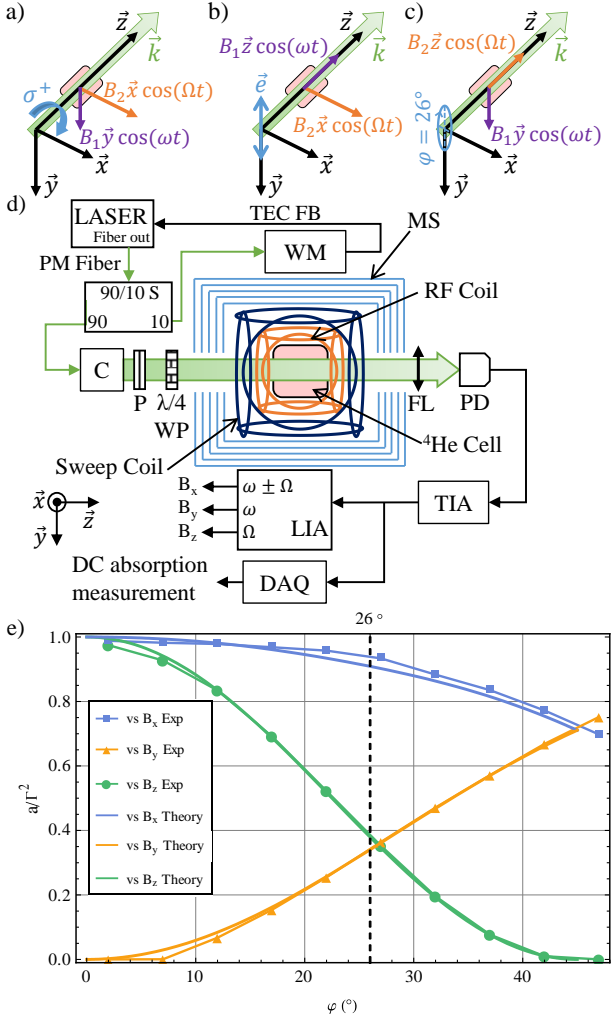

Fig. 1. (a,b,c) Schematic representation of the usual orientation-, alignment-, and elliptically-polarized-light-based PRMs configurations, respectively. (d) Experimental setup. TEC FB: TEC feedback, 90/10 S: 90/10 Splitter, WM: Wavelength-meter, MS: Magnetic shield, C: Collimator, P: Polarizer, WP: Waveplate, FL: Focusing lens, PD: Photodiode, TIA: Transimpedance amplifier, LIA: Lock-in amplifier, DAQ: DAQmx board. The LIA outputs refer the configuration of Fig. 1 (c). (e) Experimental and theoretical dependency of the amplitude ($a/\Gamma$) over width ($\Gamma$) ratio of Hanle resonances for each component of the magnetic field as a function of the pumping light ellipticity $\varphi$.

When the ellipticity is different from 0° (linear polarization) or 45° (circular polarization) one can observe Hanle resonances with respect to all components of the magnetic field.

At $\varphi = 26°$ $A$ is equal for the Hanle resonances observed by sweeping $B_y$ and $B_z$, and higher for the Hanle resonance observed by sweeping $B_x$. This ellipticity seems therefore to be a good starting point for reaching isotropic sensitivity, as it is the best compromise to have the best sensitivity to the three components of the magnetic field.

We now move to PRM scheme, by adding RF fields. We consider the setup of Fig. 1(c): an elliptically polarized pump light with $\varphi = 26°$ ellipticity propagates along $\vec{z}$, with major axis of the polarization ellipse parallel to $\vec{y}$. Two RF fields $B_1\vec{y}\cos(\omega t)$ and $B_2\vec{z}\cos(\Omega t)$ are applied, with $\omega \gg \Omega$ [17,18], with $\omega/2\pi = 40$ kHz and $\Omega/2\pi = 15$ kHz. Note that here, as shown in Fig. 1(a,b,c), unlike the usual orientation- or alignment-based PRMs [14,17,18], the RF fields are applied along the optical pumping characteristic directions: the propagation direction $\vec{k} \parallel \vec{z}$ (for the orientation) and the major axis of the polarization ellipse $\vec{y}$ (for the alignment). The reasons for this choice as well as the detailed dynamics of atoms optically pumped using elliptically polarized light and subject to several RF fields will be discussed elsewhere.

In order to find the optimal RF amplitudes we first measure the sensitivity of the PRM to each component of $\vec{B}$, as a function of the amplitudes of the two RF fields by applying a sweep of $\pm 90$ nT to each component of the field with the other set to zero. In addition to the experimental setup described above, the photodetection signal is demodulated with a Zürich MFLI lock-in amplifier, at 40 kHz to measure the $B_y$ component, at 15 kHz for the $B_z$ component and at 25 kHz ($\omega \pm \Omega$ [18]) for the $B_x$ component. The sensitivities $s_{x,y,z}$ obtained by a linear fit around the null field are shown in Fig. 2(a,b,c).

Among all the sensitivities, the largest is the one to the $B_z$ component at $\gamma B_1/\omega = 0.69$ and $\gamma B_2/\Omega = 0.57$. The sensitivity to $B_y$ reaches its maximum for $\gamma B_1/\omega = 1.02$ and $\gamma B_2/\Omega = 0.18$. Finally, the sensitivity to $B_x$ is maximum for $\gamma B_1/\omega = 0.93$ and $\gamma B_2/\Omega = 0.87$. The maximum sensitivities to $B_y$ and $B_x$ are 93 % and 90 % of the maximum to $B_z$.

A figure of merit of the overall sensitivity is $s = (s_x^2 + s_y^2 + s_z^2)^{1/2}$. Its dependence is shown in Fig. 2(d). s is maximum for $\gamma B_1/\omega = 0.93$ and $\gamma B_2/\Omega = 0.69$ (blue dot in Fig. 2(d)). However, at this maximum the three sensitivities are not equal. Experimentally, we determine that the optimal isotropic sensitivity is obtained for $\gamma B_1/\omega = 0.97$ and $\gamma B_2/\Omega = 0.76$ (green dot in Fig. 2(d)), which is in the vicinity of the isotropic condition $I_x \approx I_y \approx I_z \approx 0.33$ where $I_{x,y,z} = |s_{x,y,z}|/(|s_x| + |s_y| + |s_z|)$.

To compare this new scheme with the alignment-based PRM, we record the parametric resonance signals the two configurations sequentially on the same experimental setup (Fig. 3). For alignment-based PRM, the pumping light is linearly polarized along $\vec{y}$ ($\varphi = 0°$). The 40 kHz RF field is applied along $\vec{z}$ and the 15 kHz one along $\vec{x}$, with amplitudes so that $\gamma B_1/\omega = 0.41$ and $\gamma B_2/\Omega = 0.46$, respectively, yielding $s_x = s_z$ [18]. In the new scheme $\varphi = 26°$ and the RF amplitudes are set at optimum isotropic setting ($\gamma B_1/\omega = 0.97$ and $\gamma B_2/\Omega = 0.76$). One can see that for the two well-resolved axes of the alignment-based PRM ($B_x$ and $B_z$), the slopes are degraded of a factor 2.2 and 2.1, respectively, in the new scheme. The sensitivity to the third axis, $B_y$, is however 11 times greater.

In this new scheme, the wavelength of light has to be precisely tuned to the $^4$He $D_0$ transition. Otherwise, since the light is partially circularly polarized, it causes a vector light-shift along the propagation direction $\vec{z}$ [22], resulting in an offset and possibly increased noise on $B_z$. Another undesirable effect of such a detuning is the so-called orientation-to-alignment conversion (AOC) [23,24] which comes from the linearly polarized fraction of the pumping light. In the usual alignment-based PRM, the AOC effect do not affect the accuracy of the sensor, only possibly causing a broadening of the resonances and loss of sensitivity [24]. In the new scheme, we have observed that the AOC effect breaks the isotropy of the sensitivity and the odd-symmetry of the parametric resonance signals around the null field.

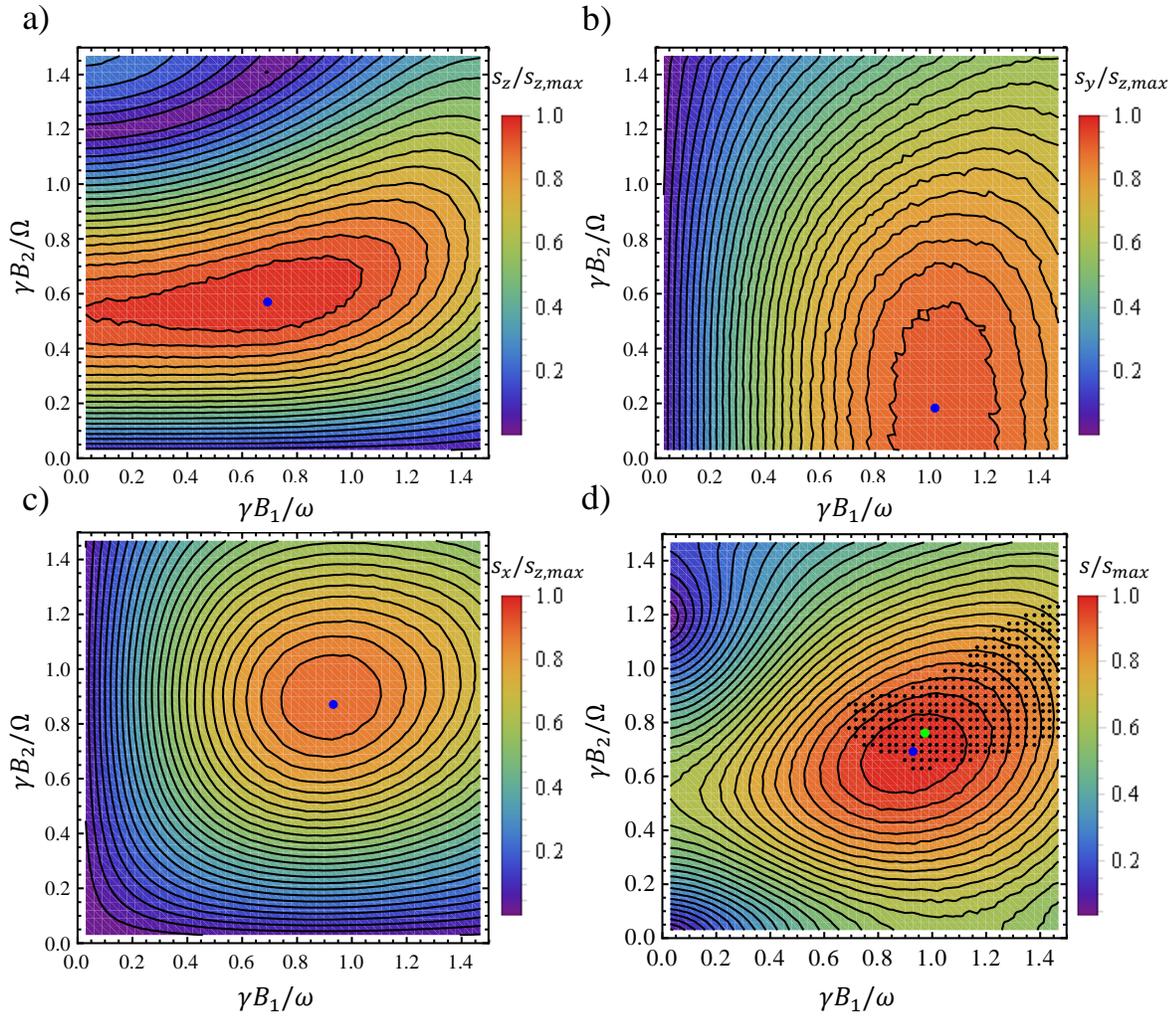

Fig. 2. (a, b, c) Experimentally measured variation of the sensitivity (normalized with the highest value reached among the three axes $s_{z,max}$) to the $B_z$, $B_y$ and $B_x$ components, respectively as a function of the RF amplitudes. These range from 16 $nT_p$ to 802.5 $nT_p$ ($\Leftrightarrow \gamma B_2/\Omega = 0.03$ to 1.5 where $\gamma = 2\pi \times 28$ Hz/nT is the $^4$He $2^3S_1$ state gyromagnetic ratio) for the slow RF field and from 42.8 $nT_p$ to 2140 $nT_p$ ($\Leftrightarrow \gamma B_1/\omega = 0.03$ to 1.5) for the fast RF field. The blue dots show the setting yielding maximum sensitivity to each component. (d) Normalized quadratic sum of the three sensitivities. The blue dot shows the setting yielding the maximum $s$. The black dotted area corresponds to the region where the sensitivity for each axis complies to $0.3 < I_x$ & $I_y$ & $I_z < 0.37$. The green dot shows the experimentally determined setting where the sensitivity is maximum and isotropic.

In conclusion, we introduced here a new PRM scheme based on elliptically polarized pumping light. With two RF fields set along the propagation and polarization directions of the pump light, it is possible to obtain a three-axis measurement of the magnetic field with isotropic sensitivity. The sensitivities of this scheme are degraded by a factor 2.2 (2.1) for the $B_x$ ($B_z$) component compared to the usual alignment-based PRMs. The sensitivity to the $B_y$ component is improved by a factor 11. With the recent improvement in the $^4$He PRMs sensitivity down to 50 fT/√Hz [25], we expect vector triaxial measurements with an isotropic sensitivity of 100 fT/√Hz.

In addition to its direct applications, isotropic sensitivity also opens interesting perspectives for building arrays of magnetometers operating in closed-loop configuration [26,27]. Indeed it avoids injecting the high magnetic noise of worst resolved axis on other axes of the neighbor magnetometers.

**Funding sources.** CEA-LETI DSYS Ph.D. funding.

**Acknowledgments**. We acknowledge technical help of W. Fourcault, cell filling by F. Alcouffe, and interesting discussions with E. Labyt, J.M. Léger, M. Le Prado, T. Jager and F. Bertrand.

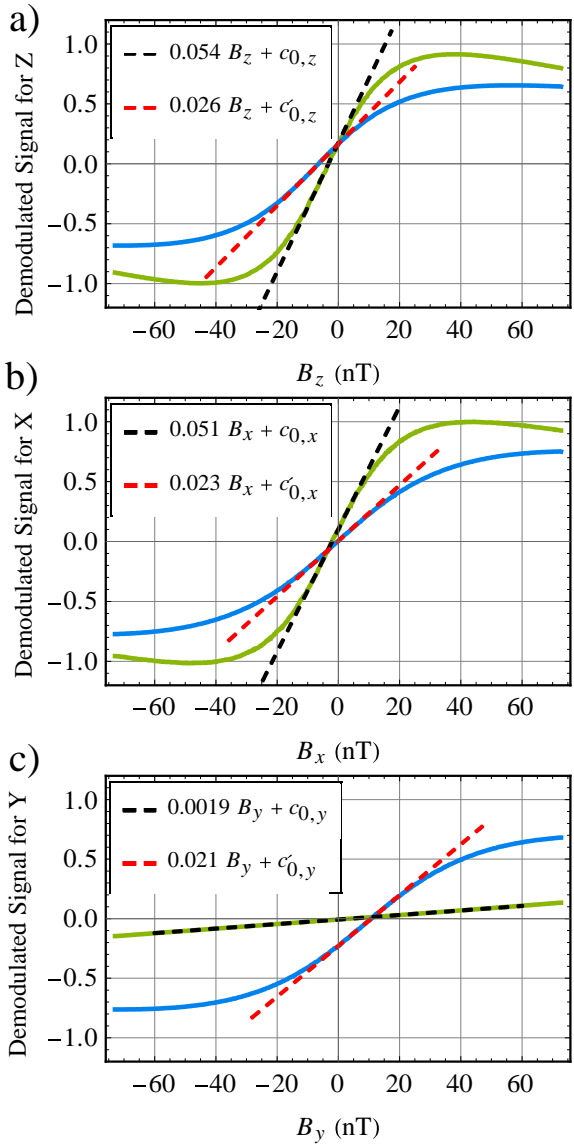

Fig. 3. Parametric resonance signals for the three components of the magnetic field and the low field linear fits, for the alignment-based standard PRM configuration (green and black dashed lines) and for the elliptically-polarized-light-based PRM configuration (blue and red dashed lines). Green curves are taken for $\gamma B_1/\omega = 0.41$ and $\gamma B_2/\Omega = 0.46$ and blue curves are acquired for $\gamma B_1/\omega = 0.97$ and $\gamma B_2/\Omega = 0.76$. All the vertical axes are normalized with the highest signal value reached among the six signals.